\documentclass[showpacs,superscriptaddress,amsfonts,amsmath,floatfix,preprint]{revtex4-1}
\usepackage[T1]{fontenc}
\usepackage[latin2]{inputenc}
\usepackage{graphicx}
\usepackage{epsfig}
\usepackage{epstopdf}

\newcommand{\I}{{\rm{i}}}

\newcommand{\tr}{\operatorname{tr}}

\newcommand{\dss}{\displaystyle}
\newcommand{\nn}{\nonumber}
\newcommand{\be}{\begin{equation}}
\newcommand{\ee}{\end{equation}}
\newcommand{\bea}{\begin{eqnarray}}
\newcommand{\eea}{\end{eqnarray}}

\newcommand{\Span}{\operatorname{span}}

\newcommand{\rank}{\operatorname{rank}}
\newcommand{\ket}[1]{| #1 \rangle}
\newcommand{\bra}[1]{\langle #1 |}
\newcommand{\braket}[2]{\langle #1 | #2 \rangle}
\newcommand{\ketbra}[2]{| #1 \rangle \langle #2 |}

\def\complex{{\mathbb{C}}}

\def\proba{{\rm I\kern -.18em P}}

\newcommand{\eg}{e.g. }
\newcommand{\ie}{i.e. }

\newcommand{\finpro}{\hfill $\Box$}

\newcommand{\Cc}{{\cal C}}

\newcommand{\Ee}{{\cal E}}

\newcommand{\Hh}{{\cal H}}

\newcommand{\Ss}{{\cal S}}

\newcommand{\Vv}{{\cal V}}
\newcommand{\Ww}{{\cal W}}

\newcommand{\Aclass}{{A\text{-cl}}}
\newcommand{\traceclass}{\rm{Mat}({\mathbb{C}},N )}
\newcommand{\opt}{{\rm{opt}}}

\begin{document}

\title {Geometric quantum discord with Bures distance}
\author{D. Spehner}
\email{Dominique.Spehner@ujf-grenoble.fr}
\affiliation{Universit\'e Grenoble 1 and CNRS, Institut Fourier  UMR5582,\\
B.P. 74, 38402 Saint Martin d'H\`eres, France}
\affiliation{Universit\'e Grenoble 1 and CNRS, Laboratoire de Physique et Mod\'elisation des
Milieux Condens\'es UMR5493, \\
B.P. 166, 38042 Grenoble, France}
\author{M. Orszag}
\affiliation{Pontificia Universidad Cat\'olica, 
Facultad de f\'{\i}sica, \\
Casilla 306, Santiago 22, Chile}  
\date{\today}

%----------------------------------------------------------------------------------

\begin{abstract}
We define a new measure of quantum correlations in bipartite quantum systems given
 by the Bures distance of the system state to the set of classical states with respect to one subsystem, 
that is, to the  states with zero quantum discord.
Our measure is a geometrical version of the quantum discord.
As the latter it quantifies the degree of non-classicality in the system.
For pure states it is identical to the geometric  measure of entanglement.    
We show that for  mixed states  it coincides with the optimal success probability of
an ambiguous quantum state discrimination task.
Moreover, the closest zero-discord states to a state $\rho$ 
are obtained in terms of the corresponding optimal measurements.
\end{abstract}

\maketitle

%%%%%%%%%%%%%%%%%%%%%%%%%%%%%%%%%%%%%%%%%%%%%%%%%%%%%%%%%%%%%%%%%%%%%
\section{Introduction} \label{sec-intro}
%%%%%%%%%%%%%%%%%%%%%%%%%%%%%%%%%%%%%%%%%%%%%%%%%%%%%%%%%%%%%%%%%%%%%

One of the basic questions in quantum information theory is to understand
how quantum correlations in composite quantum systems can be used to perform tasks that cannot 
be performed classically, or  that lead classically to  much lower efficiencies~\cite{Nielsen}.
These correlations
have  been long thought to come solely from the entanglement among the different subsystems. 
This is the case for quantum computation and communication protocols using  pure states.
   For instance, in order to offer an exponential speedup over classical computers,
a  pure-state  quantum computation must necessarily produce  multi-partite entanglement which is not 
restricted to 
blocks of qubits of fixed size as the 
problem size increases~\cite{Jozsa03}.
For composite systems in mixed states, however,  there is now increasing evidence that 
other types of quantum correlations, such as those captured by the quantum discord of 
Ollivier and Zurek~\cite{Ollivier01} and Henderson and Vedral~\cite{Henderson01},
could provide the main resource to exploit in order to outperform classical 
algorithms~\cite{Datta08,Lanyon08,Passante11,Modi_review} 
or in some quantum communication protocols~\cite{Modi_review,Madhok11,Gu2012,Dakic12}. 
The quantum discord quantifies the amount of mutual information not accessible by local measurements
on one subsystem. 
One can generate mixed states with non-zero discord but no entanglement
by preparing  locally statistical mixtures  
of nonorthogonal states, which cannot be perfectly distinguished by measurements. 
 The strongest hint so far suggesting that the discord may in certain cases quantify 
the resource responsible for quantum speedups  is provided  by the deterministic quantum
computation with one qubit  (DQC1)  of  Knill and Laflamme~\cite{Laflamme98}.
The DQC1 model leads to an exponential speedup with respect to known classical algorithms. 
   It consists of a control qubit, which remains 
 unentangled with $n$ unpolarized target qubits  at all stages of the computation. 
For other bipartitions of the $n+1$ qubits, \eg 
putting  together in one subsystem the control qubit and half of the target qubits, 
one finds in general some entanglement, but its amount is
bounded in  $n$~\cite{Datta05}. Hence,  for large system sizes, 
the total amount of bipartite entanglement is a negligible 
fraction of the maximal entanglement possible.  
On the other hand,  the DQC1 algorithm typically produces a non-zero quantum discord 
 between the control qubit and the target 
qubits~\cite{Datta08}, save in some special cases~\cite{Dakic10}. This  
 has been demonstrated experimentally in optical~\cite{Lanyon08} and 
liquid-state Nuclear Magnetic Resonance~\cite{Passante11} implementations of DQC1.
This presence of non-zero discord can be nicely interpreted by using the monogamy relation~\cite{Koashi04} 
between the discord of a 
bipartite system $AB$ and the entanglement of $B$ with its environment $E$
if $ABE$ is in a pure state~\cite{Fanchini11}.
The precise role played by the quantum discord in the DQC1 algorithm is still, however, subject to debate
(see~\cite{Modi_review} and references therein).

A mathematically appealing way to quantify quantum correlations in multi-partite systems 
is given by  the minimal distance
of the system state to a separable state~\cite{Vedral98}.
The Bures metric~\cite{Bures69,Uhlmann76} provides a nice distance $d_B$ on the convex cone 
of density matrices, which has better properties than the 
Hilbert-Schmidt distance $d_2$ from a quantum information perspective. 
In particular, $d_B$ is monotonous and Riemannian~\cite{Petz96} and its metric coincides with the 
quantum Fisher information~\cite{Caves94} playing an important role in high precision interferometry. 
As a consequence,
the minimal Bures distance to separable states satisfies all criteria of an entanglement measure~\cite{Vedral98}, 
which is not the case for the distance $d_2$. This entanglement measure 
has been widely studied in the literature~\cite{Streltsov10,Wei03,Shimony95,Rudnicki12}.
By analogy with entanglement, a geometric measure of quantum discord
has been defined by Daki\'c, Vedral, and Brukner~\cite{Dakic10} as the minimal distance of the system state
to the set of zero-discord states. This geometric quantum discord (GQD) has been evaluated explicitly 
for two qubits~\cite{Dakic10}. 
However, the aforementioned authors use the  Hilbert-Schmidt distance $d_2$, which
   leads to serious drawbacks~\cite{Piani12}.

The aim of this work is to study a similar GQD as in~\cite{Dakic10} but 
based on the Bures distance $d_B$, which seems to be a more natural choice.
This distance measure of quantum correlations has a 
clearer geometrical interpretation than other measures~\cite{Vedral98,Modi10} based on the 
relative entropy, which is not a distance on the set of density matrices.
We show that it shares many of the properties of the quantum discord.
Most importantly, as in the description of quantum correlations
using the relative entropy~\cite{Modi10,Friedland11},
our geometrical approach provides further information not contained in 
the quantum discord itself. In fact,
one can look for the closest state(s) with zero 
discord to  a given state $\rho$, and hence learn something about the ``position'' of $\rho$
 with respect to the set of zero-discord states.
The main result of this paper shows that finding the Bures-GQD and the closest zero-discord state(s) 
to $\rho$ is closely linked to 
a minimal error quantum state discrimination (QSD) problem.

The task of  discriminating  states
pertaining to a known set $\{ \rho_1, \ldots , \rho_n\}$ of density matrices $\rho_i$ with prior probabilities 
$\eta_i$ plays an important role in quantum communication and quantum cryptography.
For instance, the set $\{ \rho_1, \ldots , \rho_n\}$ can encode a message to be sent to a receiver. 
 The sender chooses at random some states among the $\rho_i$'s and gives them one by one to the receiver, who 
is required to identify them and henceforth to decode the message. 
With this goal, the receiver performs a measurement on each state given to him by the sender.
If the $\rho_i$ are non-orthogonal, they cannot be perfectly distinguished from each other by 
measurements, so that the 
amount of sent information is smaller than in the case of orthogonal states.
The best the receiver can do is to find the measurement that minimizes in some way
his probability of equivocation.
Two distinct strategies have been widely studied  in the literature 
(see the review article~\cite{Bergou_review}). 
In the first one,
the receiver seeks for a generalized measurement with $(n+1)$ outcomes, allowing him to identify
perfectly each state $\rho_i$  but such that one of the outcomes leads  to an inconclusive result 
(unambiguous QSD). The probability of occurrence of the inconclusive outcome must be
minimized. In the second strategy,
the receiver looks for a measurement with $n$ outcomes yielding the maximal success probability
$P_S= \sum_{i=1}^n \eta_i P_{i|i} $, where $P_{i|i}$ is the probability of   
the measurement outcome $i$ given that the state is $\rho_i$.
This strategy is called  minimal error (or ambiguous) QSD.
The maximal success probability $P_S^\opt$ and the optimal measurement(s) are known explicitly for $n=2$~\cite{Helstrom},
but no general solution has been found so far for more than two states (see, however,~\cite{Bae12})
except when the $\rho_i$ are related to each other by some symmetry and have equal 
probabilities $\eta_i$ (see~\cite{Bergou_review,Eldar04,Chou03} and references therein). 
However, several upper bounds on $P_S^\opt$ are known~\cite{Qiu10}
 and the discrimination task can be  solved efficiently numerically~\cite{Helstrom82,Jezek12}.    
Let us also stress that unambiguous and ambiguous QSD have been implemented 
experimentally for pure states~\cite{Clarke01} and, more recently, for mixed states~\cite{Bergou04}, by  using 
polarized light.

   Let  $\rho$ be any state of a bipartite system
with a finite-dimensional Hilbert space.
We will prove in what follows that the Bures-GQD of $\rho$ 
is equal to the maximal success probability $P_S^\opt$ in the ambiguous QSD
of a family of states $\{ \rho_i\}$ and prior probabilities $\{ \eta_i\}$ depending on $\rho$. Moreover, 
the closest zero-discord states to $\rho$ are given in terms
of the corresponding optimal von Neumann measurement(s). The number of states $\rho_i$ to discriminate
is equal to the dimension of the Hilbert space of the measured subsystem.
When this subsystem is a qubit, the discrimination task involves only 
two states and can be solved exactly~\cite{Bergou_review,Helstrom}: $P_S^\opt$ and the
optimal von Neumann projectors are given in terms of the eigenvalues and eigenvectors
of the hermitian matrix $\Lambda=\eta_0 \rho_0 - \eta_1 \rho_1$. 
In a companion paper~\cite{companion_paper}, we use this approach to 
derive an explicit formula for the 
Bures-GDQ of a family of two-qubits states (states with maximally mixed marginals) and
determine the corresponding closest zero-discord states.

This article is organized as follows. The 
definitions of the quantum discords and of the Bures distance are given in Sect.~\ref{sec-definitions}, 
together with their main properties. 
In Sect.~\ref{sec_pure_states}, we show that the Bures-GQD 
of a pure state coincides with the geometric measure of entanglement and is simply related 
to the highest Schmidt coefficient. We explain this fact by noting that the closest zero-discord
states to a pure state are convex combinations of orthogonal pure product states.
The link between the minimal Bures distance to the set of zero-discord states 
and ambiguous QSD
is explained and proved in the Sect.~\ref{sec-mixed_state}. 
The last section contains some conclusive remarks and perspectives.
The appendix contains a technical proof of an intuitively obvious fact in QSD. 

%%%%%%%%%%%%%%%%%%%%%%%%%%%%%%%%%%%%%%%%%%%%%%%%%%%%%%%%%%%%%%%%%%%%%%
\section{Definitions of the  quantum discords and Bures distance} \label{sec-definitions}
\subsection{The quantum discord and the set of $A$-classical states    }
%%%%%%%%%%%%%%%%%%%%%%%%%%%%%%%%%%%%%%%%%%%%%%%%%%%%%%%%%%%%%%%%%%%%%

In this work we consider a bipartite quantum system $AB$ with Hilbert space $\Hh=\Hh_A \otimes \Hh_B$,
the spaces $\Hh_A$ and $\Hh_B$  of the subsystems $A$ and $B$ having arbitrary finite dimensions
$n_A$ and $n_B$. The states of $AB$ are given by density matrices $\rho$ on $\Hh$
(\ie Hermitian positive $N\times N$ matrices $\rho \in \traceclass$ with unit trace $\tr(\rho)=1$, 
with $N =n_A n_B$). 
The reduced states of  $A$ and $B$ are defined by partial tracing $\rho$ 
over the other subsystem. They  are denoted by
 $\rho_A = \tr_{B} (\rho)$ and $\rho_B=\tr_A(\rho)$.

Let  us first recall the definition of the quantum discord~\cite{Ollivier01,Henderson01}.
The total correlations of the bipartite system in the state $\rho$ are described by the mutual information 
$I_{A:B}(\rho )= S(\rho_A) +  S(\rho_B)-S(\rho)$,
where $S(\cdot )$ stands for the von Neumann entropy.  
The amount $J_{B | A} (\rho)$ of classical correlations is given by the maximal reduction  of entropy
of the subsystem $B$ after a von Neumann measurement on $A$. Such a measurement is described by
an orthogonal family $\{ \pi_i^A\}$ of projectors acting on $\Hh_A$ 
(\ie by self-adjoint operators $\pi_i^A$  on $\Hh_A$ satisfying $\pi_i^A \pi_j^A = \delta_{ij} \pi_i^A$). Hence
$J_{B | A} (\rho) =  \max_{ \{ \pi_i^A \} } \{  S( \rho_B) - \sum_i q_i S ( \rho_{B|i}  ) \}$, 
where the maximum is over all von Neumann measurements  $\{ \pi_i^A\}$, 
$q_i = \tr ( \pi_i^A \otimes 1\, \rho )$ is the probability
of the measurement outcome $i$, and 
$\rho_{B|i} = q_i^{-1} \tr_A (  \pi_i^A \otimes 1 \,\rho )$ is the corresponding post-measurement conditional
state of $B$. 
The quantum discord is by definition the difference  $\delta_{A} ( \rho) = I_{A:B}(\rho )- J_{B | A} (\rho)$
between the total and classical correlations. It measures the amount of mutual information which is not accessible 
by local measurements on the subsystem $A$. 
Note that it is asymmetric under the exchange $A \leftrightarrow B$.
It can be shown~\cite{Datta11} that $\delta_A (\rho)\geq 0$ for any $\rho$. Moreover, 
$\delta_A(\sigma_{\Aclass})=0$ if and only if 
\be \label{eq-A-classical_state}
\sigma_{\Aclass} 
=
\sum_{i=1}^{n_A}  q_i \ketbra{\alpha_i}{\alpha_i} \otimes \sigma_{B|i} \;,
\ee
where $\{ \ket{\alpha_i} \}_{i=1}^{n_A}$ is an orthonormal basis of $\Hh_A$, 
$\sigma_{B|i}$ are some (arbitrary)
states of $B$ depending on the index $i$, and $q_i \geq 0$ are some probabilities, $\sum_i q_i = 1$.
The fact  that  $\delta_A(\sigma_{\Aclass})=0$ follows directly from  
$I_{A:B}(\sigma_\Aclass )=S (\tr_A ( \sigma_\Aclass ) ) - \sum_i q_i S (\sigma_{B|i}) \leq J_{B|A} (\sigma_\Aclass)$ and
from the non-negativity of the quantum discord. 
For a bipartite system in the  state $\sigma_{\Aclass}$, 
the subsystem $A$ is in one of the orthogonal states $\ket{\alpha_i}$ with probability $q_i$, 
hence $A$ behaves as a classical system. 
For this reason, we will call {\it $A$-classical states}
the zero-discord states of the form (\ref{eq-A-classical_state}). 
In the literature they are often referred to as the ``classical-quantum'' states.
We denote by $\Cc_A$ the set of all $A$-classical states.
By using the spectral decompositions of the $\sigma_{B|i}$, any $A$-classical
state $\sigma_{\Aclass} \in \Cc_A$ can be decomposed as
\be \label{eq-A-classical_state_bis}
\sigma_{\Aclass} 
=
\sum_{i=1}^{n_A} \sum_{j=1}^{n_B} q_{ij} \ketbra{\alpha_i}{\alpha_i} \otimes \ketbra{\beta_{j|i}}{\beta_{j|i}}
\ee
where, for any fixed $i$, $\{ \ket{\beta_{j|i}} \}_{j=1}^{n_B}$ is an orthonormal basis of
$\Hh_B$,  and $q_{ij} \geq 0$, $\sum_{ij} q_{ij} = 1$
(note that the 
$\ket{\beta_{j|i}}$ need not be orthogonal for distinct $i$'s). 
One defines similarly  the set
$\Cc_B$ of $B$-classical states, which are the states with zero quantum discord when the subsystem $B$ 
is measured. A state which is both $A$- and $B$-classical possesses an
eigenbasis  $\{ \ket{\alpha_i} \otimes \ket{\beta_j}\}_{i=1,j=1}^{n_A,n_B}$ of product vectors. It
is fully classical, in the sense that a quantum system in this state can be
``simulated'' by a classical apparatus being in the state 
$(i,j)$ with probability $q_{ij}$.
 
Let us point out that $\Cc_A$, $\Cc_B$, and the set of classical states $\Cc$ are not convex. Their convex 
hull is the set $\Ss$ of separable states. A state $\sigma_{\rm sep}$ is separable if it admits
a convex decomposition 
$\sigma_{\rm sep}= \sum_{m} q_m \ketbra{\phi_m}{\phi_m} \otimes \ketbra{\psi_m}{\psi_m}$, where
$\{ \ket{\phi_m}\} $ and $\{ \ket{\psi_m}\}$ are (not necessarily orthogonal) families of 
unit vectors in $\Hh_A$ and $\Hh_B$ and $q_m \geq 0$, $\sum_m  q_m =1$.
For pure states, $A$-classical, 
$B$-classical, classical and separable 
states all coincide. Actually, according to  (\ref{eq-A-classical_state_bis}) the 
pure $A$-classical (and, similarly, the pure $B$-classical) states are product states.

%%%%%%%%%%%%%%%%%%%%%%%%%%%%%%%%%%%%%%%%%%%%%%%%%%%%%%%%%%
\subsection{Distance measures of quantum correlations with the Bures distance}
%%%%%%%%%%%%%%%%%%%%%%%%%%%%%%%%%%%%%%%%%%%%%%%%%%%%%%%%%%%

The geometric quantum discord (GQD) of a state $\rho$ of $AB$
has been  defined in~\cite{Dakic10} as the square distance of 
$\rho$ to the set $\Cc_A$ of $A$-classical states,
\begin{equation} \label{eq-def_geo_discord}
D_A^{(2)} (\rho) = d_2 ( \rho, \Cc_A)^2 = \min_{\sigma_\Aclass \in \Cc_A} d_2 ( \rho,\sigma_\Aclass )^2
\end{equation} 
where  $d_2(\rho,\sigma) = ( \tr [ (\rho-\sigma)^2] )^{1/2}$ is the Hilbert-Schmidt distance.
Instead of taking this distance, we use in this article the Bures distance    
\be \label{eq-Bures_distance}
d_B ( \rho, \sigma) = \Bigl[ 2 \bigl( 1 - \sqrt{F(\rho,\sigma)} \bigr) \Bigr]^{\frac{1}{2}}
\ee
where  $\rho$ and $\sigma$ are two density matrices and $F(\rho,\sigma)$ is their 
fidelity~\cite{Nielsen,Uhlmann76,Jozsa94}, 
\be \label{eq-fidelity}
F(\rho,\sigma) 
= \| \sqrt{\rho} \sqrt{\sigma} \|_1^2 
 = \bigl[ \tr \bigl(  [ \sqrt{\sigma} \rho \sqrt{\sigma} ]^{1/2} \bigr) \bigr]^2 \;.
\ee

It is known that (\ref{eq-Bures_distance}) defines a Riemannian distance on the convex cone 
$\Ee \subset \traceclass$ of all density matrices of $AB$. Its metric 
 is equal to the Fubini-Study metric for pure states and coincides (apart from a numerical factor) with 
the quantum Fisher information which plays an important role in 
quantum metrology~\cite{Caves94}. 
Moreover, $d_B$ satisfies the 
following properties~\cite{Jozsa94,Nielsen}: for any  $\rho,\sigma,\rho_1, \rho_2, \sigma_1$, and 
$\sigma_2 \in \Ee$,
\\
(i) {\it joint convexity of the square distance:} if $\eta_1,\eta_2 \geq 0$ and  $\eta_1+\eta_2 = 1$, then
$d_B(\eta_1 \rho_1 + \eta_2 \rho_2 ,\eta_1 \sigma_1 + \eta_2 \sigma_2 )^2 \leq  \eta_1 d_B (\rho_1,\sigma_1)^2 
+ \eta_2 d (\rho_2,\sigma_2)^2$;\\
(ii) {\it $d_B$ is monotonous under the action of completely positive 
trace-preserving maps ${\cal T}$ from $\traceclass$ into itself:} 
for any such $\cal{T}$, $d_B ( {\cal T} \rho, {\cal T} \sigma ) \leq d_B ( \rho , \sigma)$.

Property (ii) implies that
 $d_B$ is invariant under unitary conjugations: if $U$ is a unitary operator on $\Hh$, then
$d_B ( U \rho U^\dagger, U \sigma U^\dagger)= d_B (\rho,\sigma)$. 
Note that the Hilbert-Schmidt distance $d_2$ is also unitary invariant 
 but  fails to satisfy (ii)
   (a simple counter-example can be found in~\cite{Ozawa00}).
The monotonous Riemannian distances on $\Ee$ have been  classified by Petz~\cite{Petz96}. 
The Bures distance can be used to bound from below and above the
trace distance $d_1(\rho,\sigma) = \tr ( | \rho-\sigma|)$ as follows~\cite{Nielsen},
\begin{equation}
d_B (\rho,\sigma)^2 \leq d_1(\rho,\sigma) 
 \leq 
  \Bigl[ 1 - \Bigl( 1 -\frac{1}{2} d_B(\rho,\sigma)^2 \Bigr)^2 \Bigr]^\frac{1}{2}
\;.
\end{equation}
   For good reviews on the Uhlmann fidelity and Bures distance, see the book of 
Nielsen and Chuang~\cite{Nielsen} and the nice introduction of the article~\cite{Sommers03}
devoted to the estimation of the Bures volume of $\Ee$.

We define the GQD as    
\begin{equation} \label{eq-max_fidelity} 
D_A ( \rho) = d_B(\rho, \Cc_A)^2 = 2  ( 1 - \sqrt{ F_A (\rho )} ) 
\quad , \quad  F_A(\rho)= \max_{\sigma_\Aclass \in \Cc_A} F (\rho, \sigma_\Aclass)\;.
\end{equation}
The unitary invariance of $d_B$ and $d_2$  implies  
that $D_A$ and $D_A^{(2)}$ are invariant under conjugations by local unitaries,
$\rho \mapsto  U_A \otimes U_B \rho\, U_A^\dagger \otimes U_B^\dagger$, since such transformations leave
$\Cc_A$ invariant. By property (ii),
$D_A$ is monotonous under local operations involving von Neumann measurements on $A$ and generalized
measurements on $B$.   

By analogy with (\ref{eq-max_fidelity}), one can define two other geometrical measures of quantum correlations: 
the square distance 
 to the set of classical states $\Cc$ and
the  geometric measure of entanglement,
\begin{equation} \label{eq_def_D_and_E}
D (\rho) = d_B ( \rho, \Cc)^2  =2(1-\sqrt{F_\Cc (\rho)}) \quad , \quad 
E (\rho) = d_B ( \rho, \Ss)^2 =2(1-\sqrt{F_\Ss (\rho)})\;,
\end{equation}
where $F_\Cc (\rho)$ is  the maximal fidelity between $\rho$ and a classical state
$\sigma_{\rm cl} \in \Cc$ and $F_\Ss (\rho)$ the maximal fidelity between $\rho$ and a separable state 
$\sigma_{\rm sep} \in \Ss$. 
   The first measure $D$ is a geometrical analogue of the measurement-induced 
disturbance (MID)~\cite{Luo08a}, which has up to our knowledge not been studied so far 
(however, an analogue of the MID  based on the relative entropy has been introduced in \cite{Modi10}). 
The second measure $E$ satisfies all criteria of an 
entanglement measure~\cite{Vedral98} (in particular, it is monotonous under 
local operations and classical communication by the property (ii)) and has 
been studied in~\cite{Vedral98,Streltsov10,Rudnicki12}. It
is closely related to other entanglement measures~\cite{Wei03,Shimony95} defined via a convex roof construction
thanks to the identity~\cite{Streltsov10}
\begin{equation} \label{eq-fidelity_as_convex_roof}
F_\Ss (\rho)  = \max_{\{ p_m\} , \{ \ket{\Psi_m}\}} \sum_m p_m 
F_\Ss ( \ket{\Psi_m}  )\quad , \quad 
F_\Ss ( \ket{\Psi_m} ) =  \max_{\sigma_{\rm{sep}} \in \Ss} F ( \ketbra{\Psi_m}{\Psi_m}, \sigma_{\rm sep})\;,
\end{equation}
where the maximum is over all  pure state decompositions  
$\rho=\sum_m p_m \ketbra{\Psi_m}{\Psi_m}$ of $\rho$ (with  $\| \Psi_m\|=1$ and $p_m\geq 0$,
$\sum p_m =1$). 
The measure $E$ is a geometrical analogue of 
the entanglement of formation~\cite{Bennett96}. The latter is defined via a convex roof 
construction from the von Neumann entropy of the reduced state, 
\begin{eqnarray}
\nn
E_{\rm{EoF}} (\rho) & = & \min_{\{ p_m \} ,\{ \ket{\Psi_m}\}} \sum_m p_m  E_{\rm{EoF}} (\ket{\Psi_m})\;,
\\ & & \hspace*{2cm} 
E_{\rm{EoF}} (\ket{\Psi_m})=S ( \tr_A (\ketbra{\Psi_m}{\Psi_m}) )=S ( \tr_B (\ketbra{\Psi_m}{\Psi_m}) )\;. 
\end{eqnarray}

Since 
$\Cc \subset \Cc_A \subset \Ss$, the three distances are ordered as 
\be
E(\rho) \leq D_A (\rho) \leq D(\rho)\,.
\ee 
This ordering of quantum correlations is a nice feature of the geometric 
measures. In contrast, the 
entanglement of formation $E_{\rm{EoF}} (\rho)$ 
can be larger or smaller than the quantum discord $\delta_A (\rho)$~\cite{Orszag12a,Orszag12b}. 

%%%%%%%%%%%%%%%%%%%%%%%%%%%%%%%%%%%%%%%%%%%%%%%%%%%%%%%%%%%%%%%%%%%%%%%%%%%%%
\section{The Bures geometric quantum discord of pure states} \label{sec_pure_states}
%%%%%%%%%%%%%%%%%%%%%%%%%%%%%%%%%%%%%%%%%%%%%%%%%%%%%%%%%%%%%%%%%%%%%%%%%%%%%
 
We first 
restrict our attention to pure 
states, for which one can obtain a simple formula for
$D_A$ in terms of the Schmidt
coefficients $\mu_i$. We recall that any pure state 
$\ket{\Psi} \in \Hh_A \otimes \Hh_B$ admits a Schmidt decomposition
\be \label{eq-Schmidt_decomposition}
\ket{\Psi} = \sum_{i=1}^{n} \sqrt{\mu_i} \ket{\varphi_i} \otimes \ket{\chi_i}
\ee
where $n =\min \{ n_A,n_B\}$ and $\{ \ket{\varphi_i} \}_{i=1}^{n_A}$ 
(respectively $\{ \ket{\chi_j} \}_{j=1}^{n_B}$) is
an orthonormal basis of $\Hh_A$ ($\Hh_B$). If the $\mu_i$ are non-degenerate,
the decomposition (\ref{eq-Schmidt_decomposition}) is unique,
the $\mu_i$ and $ \ket{\varphi_i} $ (respectively $\ket{\chi_j}$) being the eigenvalues and
 eigenvectors of the reduced state
$(\ketbra{\Psi}{\Psi})_A$ (respectively $(\ketbra{\Psi}{\Psi})_B$). 
Note that $\mu_i \geq 0$ and $\sum_i \mu_i = \| \Psi\|^2  = 1$.

\vspace{2mm}

\noindent {\bf Theorem 1.} {\it If $\rho_\Psi = \ketbra{\Psi}{\Psi}$ is a pure state, then  
\be \label{eq-equality_distances}
D_A (\rho_\Psi ) = D (\rho_\Psi )  = E (\rho_\Psi )  
 =  2 ( 1 - \sqrt{\mu_{\rm max}} ) \;,
\ee
where $\mu_{\rm max}$ is the largest Schmidt 
eigenvalue $\mu_i$. If this maximal eigenvalue  is non-degenerate, 
the closest $A$-classical (respectively classical, separable) state to $\rho_\Psi$ is the pure product 
state 
$\sigma  = \ketbra{\varphi_{{\rm max}} \otimes \chi_{{\rm max}}}{\varphi_{{\rm max}} \otimes \chi_{{\rm max}}}$,
where $\ket{\varphi_{{\rm max}}}$ and $\ket{ \chi_{{\rm max}}}$ are the eigenvectors corresponding to $\mu_{\rm max}$
 in the decomposition (\ref{eq-Schmidt_decomposition}).
If $\mu_{\rm max}$ is $r$-fold degenerate, say 
$\mu_{\rm max}=\mu_1=\cdots = \mu_r> \mu_{r+1}, \ldots , \mu_{n}$, then infinitely many
$A$-classical (respectively
classical, separable) states $\sigma$ minimize the distance $d_B(\rho_\Psi,\sigma )$.
These closest states $\sigma$ are convex combinations of the orthogonal pure product states 
$\ketbra{\alpha_{l} \otimes \beta_{l}}{\alpha_{l} \otimes \beta_{l}}$, $l=1,\ldots, r$,
with 
$\ket{\alpha_{l}} = \sum_{i=1}^r u_{il} \ket{\varphi_i}$ and
$\ket{\beta_{l} }= \sum_{i=1}^r u_{il}^\ast \ket{\chi_i}$, where $(u_{il})_{i,l=1}^r$ is an
 arbitrary $r\times r$ unitary matrix
and $\ket{\varphi_i}$ and $\ket{\chi_i}$ are some 
eigenvectors in the  decomposition (\ref{eq-Schmidt_decomposition}).}

\vspace{3mm}

The expression (\ref{eq-equality_distances}) of the geometric measure of entanglement
$E(\rho_\Psi)$ is basically known in the literature~\cite{Shimony95,Wei03}. The closest separable states 
to pure and mixed states  have been investigated in~\cite{Streltsov10}.
By inspection of (\ref{eq-Schmidt_decomposition}) and (\ref{eq-equality_distances}),
$D_A (\rho_\Psi )=0$ 
if and only if $\ket{\Psi}$ is a product state,
in agreement with the fact that  $A$-classical
pure states are  product states (the same holds for the other quantum correlation measures 
$D$ and $E$).  
Moreover, from the inequality $\mu_{\rm max} \geq 1/n$ (following from
$\sum_{i=1}^n \mu_i = 1$) one deduces that
$D_A (\rho_\Psi ) \leq 2  (  1 - 1/\sqrt{n} )$. The maximal value of $D_A$ is reached when 
$\mu_i = 1/n$ for any $i$, that is, for the maximally entangled states
(recall that such states are the pure states with   reduced states 
$(\rho_\Psi)_A$ and $(\rho_\Psi)_B$ having a maximal entropy 
$S ((\rho_\Psi)_A) = - \sum_{i=1}^{n} \mu_i \ln \mu_i = \ln ( n )$).

Note that when $\mu_{\rm max}$ is $r$-fold degenerate, the $r$ vectors 
$| \alpha_l \rangle$ (respectively $| \beta_l \rangle$) are 
orthonormal eigenvectors of $[ \rho_\Psi]_A$ (respectively $[\rho_\Psi]_B$) with eigenvalue $\mu_{\rm max}$. 
One then obtains another Schmidt decomposition of $| \Psi\rangle$ by replacing  in (\ref{eq-Schmidt_decomposition}) the $r$ 
eigenvectors $| \varphi_i \rangle$ and $| \chi_i \rangle$ with eigenvalue $\mu_{\rm max}$ by $| \alpha_l \rangle$ and $| \beta_l \rangle$.

Remarkably, the maximally entangled states are the pure states admitting the largest family of closest
separable states (this family is a  $(n^2+n-2)$ real-parameter submanifold of $\Ee$). 
For instance, in the case of two qubits (\ie for $n_A = n_B = n=2$),  
the Bell states $\ket{\Phi^\pm} = ( \ket{00} \pm \ket{11} )/\sqrt{2}$ admit as closest separable states 
the classical states
\begin{equation}
\sigma_\pm 
 = 
  \sum_{l=0,1} q_l \ketbra{\alpha_l}{\alpha_l} \otimes \ketbra{\beta_l}{\beta_l}
\;\;,\;\;
\ket{\alpha_l} =  u_{0l} \ket{0} + u_{1l} \ket{1}
\;\;,\;\;
\ket{\beta_l} =  u_{0l}^\ast \ket{0} \pm u_{1l}^\ast \ket{1} 
\end{equation}
with $u_{0l}^\ast u_{0m} + u_{1l}^\ast u_{1m} = 
\delta_{ml}$ and $q_l\geq 0$, $q_0+q_1=1$.
Interestingly, typical decoherence processes such as pure phase dephasing
transform  $\rho_{\Phi^{\pm}}$ into one 
of its closest separable state $(\ketbra{00}{00} + \ketbra{11}{11})/2$
 at times $t \gg t_{\rm dec}$, where $t_{\rm dec}$ is the decoherence time. Slower 
relaxation processes modifying the populations in the states $\ket{00}$ and $\ket{11}$
do not further increase the distance to the initial state $\rho_{\Phi^{\pm}}$.  
The situation is different for a partially entangled state 
$\ket{\Psi} = \sqrt{\mu_0} \ket{00} + \sqrt{\mu_1} \ket{11}$ with $\mu_1 > \mu_0$: then 
the closest separable state  is the pure state $\ket{11}$, but $\ket{\Psi}$ evolves asymptotically 
to a statistical mixture of $\ket{00}$ and 
$\ket{11}$  when the  qubits are coupled e.g. to thermal baths at positive temperatures.

\vspace{2mm}

\noindent {\it Proof of Theorem~1.}   
For a pure state $\rho_\Psi$, the fidelity reads 
$F (\rho_\Psi, \sigma_\Aclass ) = \bra{\Psi} \sigma_\Aclass \ket{\Psi}$. 
Replacing $\sigma_\Aclass$ in (\ref{eq-max_fidelity}) by the right-hand
side of (\ref{eq-A-classical_state_bis}) we get
\be \label{eq-max_fidelity_pure_state}
F_A( \rho_\Psi ) 
 =  \max_{ \{ \ket{{\alpha}_i} \}, \{ \ket{{\beta}_{j|i}} \}, \{ q_{ij} \} }
 \Bigl\{ \sum_{ij} q_{ij} | \braket{{\alpha}_i \otimes {\beta}_{j|i}}{\Psi} |^2 \Bigr\}
 =  
\max_{\| \alpha\| = \| \beta\|=1}  | \braket{\alpha \otimes \beta}{\Psi} |^2  
\ee
where we have used $\sum_{ij} q_{ij} =1$.
Thanks to the Cauchy-Schwarz inequality, for any normalized vectors
 $\ket{\alpha}\in \Hh_A$ and $\ket{\beta}\in \Hh_B$ one has 
\begin{eqnarray} \label{eq-bound_by_mu_max}
\nn
|\braket{\alpha \otimes \beta}{\Psi} | 
& = &  \Bigl|  \sum_{i=1}^n \sqrt{\mu_i}  \braket{\alpha}{\varphi_i} \braket{\beta}{\chi_i}  \Bigr| 
\\
& \leq & 
\sum_{i=1}^n \sqrt{\mu_i} \bigl| \braket{\alpha}{\varphi_i} \braket{\beta}{\chi_i} \bigr|
\leq 
\sqrt{\mu_{\rm max}}  \sum_{i=1}^n  \bigl| \braket{\alpha}{\varphi_i} \braket{\beta}{\chi_i} \bigr| 
 \\ \label{eq-bound_by_mu_max2}
& \leq & 
\sqrt{\mu_{\rm max}} \Bigl[ \sum_{i=1}^n | \braket{\alpha}{\varphi_i} |^2 \Bigr]^{1/2} 
   \Bigl[ \sum_{i=1}^n | \braket{\beta}{\chi_i} |^2 \Bigr]^{1/2}
    \leq \sqrt{\mu_{\rm max}} \;.
\end{eqnarray}

Let us first assume that $\mu_1=\mu_{\rm max}>\mu_2, \ldots , \mu_n$.
Then $|\braket{\alpha \otimes \beta}{\Psi} | = \sqrt{\mu_{\rm max}}$ if and only if 
$\ket{\alpha} = \ket{\varphi_1}$ and $\ket{\beta} = \ket{\chi_1}$ up to irrelevant phase factors. Thus 
the maximal fidelity $F_A( \rho_\Psi )$ between $\rho_\Psi$ and an 
$A$-classical state is simply given by the largest Schmidt eigenvalue $\mu_{\rm max}$. 
Moreover, the maximum in the second member of Eq.(\ref{eq-max_fidelity_pure_state})  
is reached when a single $q_{ij}$ is non-vanishing, say  
$q_{ij} = \delta_{i 1} \delta_{j 1}$, and $\ket{{\alpha}_{1}}=\ket{\varphi_1}$, 
 $\ket{{\beta}_{1|1}}=\ket{\chi_1}$.
This means that
the closest $A$-classical state to $\rho_\Psi$ is the pure product state 
$\ketbra{\varphi_{1} \otimes \chi_{1} }{\varphi_{1} \otimes \chi_{1} }$.
Since this is a classical state, one has
$F_\Cc (\rho_\Psi)=F_A( \rho_\Psi ) = \mu_{\rm max}$. One shows similarly that $F_\Ss (\rho_\Psi)=\mu_{\rm max}$. Then 
(\ref{eq-equality_distances}) follows from the definitions (\ref{eq-max_fidelity}) and
(\ref{eq_def_D_and_E}) of $D_A$, $D$, and $E$.

More generally, let  
$\mu_1=\cdots = \mu_r=\mu_{\rm max}> \mu_{r+1}, \ldots , \mu_{n}$.
We need to show that all inequalities in  (\ref{eq-bound_by_mu_max}) and (\ref{eq-bound_by_mu_max2})
are equalities
for appropriately  chosen normalized vectors $\ket{\alpha}$ and $\ket{\beta}$.
The first  inequality in  (\ref{eq-bound_by_mu_max}) is an equality if and only if 
$\arg ( \braket{\alpha}{\varphi_i} \braket{\beta}{\chi_i}) =\theta$ 
is independent of $i$. The second inequality in  (\ref{eq-bound_by_mu_max})  is an equality if and only if
$\ket{\alpha} $ belongs to $V_{\rm max} = \Span \{ \ket{\varphi_i} \}_{i=1}^r$
or $\ket{\beta}$ belongs to $W_{\rm max} = \Span \{ \ket{\chi_i} \}_{i=1}^r$. 
The Cauchy-Schwarz inequality in (\ref{eq-bound_by_mu_max2}) is an equality if and only if
$| \braket{\alpha}{\varphi_i} | =  \lambda | \braket{\beta}{\chi_i} |$ for all $i$, with $\lambda\geq 0$. 
Finally, the last inequality in (\ref{eq-bound_by_mu_max2}) is an equality if and only if
both sums inside the square brackets are equal to unity, \ie
$\ket{\alpha} \in \Span \{ \ket{\varphi_i}\}_{i=1}^n$ and $\ket{\beta} \in \Span \{ \ket{\chi_i}\}_{i=1}^n$
(this holds trivially if $n_A=n_B=n$). Putting all conditions together, we obtain
$\ket{\alpha} \in V_{\rm max}$, $\ket{\beta} \in W_{\rm max}$, and 
$ \braket{\beta}{\chi_i} = e^{\I \theta} \braket{\varphi_i}{\alpha}$ for $i=1,\ldots,r$.
Therefore,
from any orthonormal family $\{ \ket{\alpha_{l}} \}_{l=1}^r$ of $V_{\rm max}$  one can  construct  
$r$ orthogonal vectors $\ket{\alpha_{l} \otimes \beta_{l}}$ satisfying
$|\braket{\alpha_{l} \otimes \beta_{l}}{\Psi} | = \sqrt{\mu_{\rm max}}$ for all $l=1,\ldots, r$, with   
$ \braket{\beta_l}{\chi_i} = \braket{\varphi_i}{\alpha_l}$.  
The probabilities $\{ q_{ij}\}$ maximizing the sum  inside the brackets in (\ref{eq-max_fidelity_pure_state})
are given by $q_{ij}=q_i$  if $i=j \leq r$ and zero otherwise,
where 
$\{ q_l\}_{l=1}^r$ is an arbitrary set of probabilities.
The corresponding $A$-classical states with maximal
fidelities $F(\rho_\Psi,\sigma)$ are the classical states
$\sigma 
 = 
  \sum_{l=1}^r q_l 
   \ketbra{\alpha_{l} \otimes \beta_{l}}{\alpha_{l} \otimes \beta_{l}}
$.
\finpro

\vspace{2mm}

The equality between the correlation measures $D_A$, $D$, and $E$  is a consequence of the 
fact that the closest states to $\rho_\Psi$ are classical states. 
Such an equality is reminiscent from the equality between the entanglement of formation $E_{\rm EoF}$
and the quantum discord $\delta_A$ for pure states.
Let us notice that it does {\it not} hold for the Hilbert-Schmidt distance, for which the closest $A$-classical 
state to a pure state is in general a mixed state. Actually, one infers from the expression
\be \label{eq-d2_is_bad}
d_2 ( \rho_\Psi, \sigma_{\Aclass} )^2 = \tr[ ( \ketbra{\Psi}{\Psi} - \sigma_{\Aclass} )^2 ]
= 1 - 2 F( \rho_\Psi, \sigma_{\Aclass} ) + \tr ( \sigma_{\Aclass}^2 ) 
\ee
that the closest $A$-classical state results from a competition between the  maximization of 
the fidelity $F( \rho_\Psi, \sigma_{\Aclass} )$ and the minimization of the trace $\tr ( \sigma_{\Aclass}^2)$,
which is maximum for pure states. For instance, one can show~\cite{companion_paper} that
the closest $A$-classical states to the Bell states $\ket{\Phi^\pm}$ for $d_2$
are mixed two-qubit states.
The validity of Theorem 1 is one of the major advantage of the Bures-GQD over
the Hilbert-Schmidt-GQD.

%%%%%%%%%%%%%%%%%%%%%%%%%%%%%%%%%%%%%%%%%%%%%%%%%%%%%%%%%%%%%%%%%%%%%%%%%%%%%
\section{The Bures geometric quantum discord of mixed states}\label{sec-mixed_state}
\subsection{Link with minimal error quantum state discrimination} \label{eq-main_result}
%%%%%%%%%%%%%%%%%%%%%%%%%%%%%%%%%%%%%%%%%%%%%%%%%%%%%%%%%%%%%%%%%%%%%%%%%%%%%  

The determination of $D_A (\rho)$ is much more involved for mixed states  
than for pure states. We show in this section that this problem is related  to 
ambiguous QSD.
As it has been recalled in the introduction, in ambiguous QSD 
a state  $\rho_i$ drawn from a known family
$\{ \rho_i \}_{i=1}^{n_A}$ with prior probabilities $\{ \eta_i \}_{i=1}^{n_A}$ is sent to a receiver. 
The task of the latter is to determine which state he has received
with a maximal probability of success. 
To do so, he performs a generalized measurement
and concludes that the state is $\rho_j$ when his measurement result is $j$.
The generalized measurement  is given by a family of positive operators  $M_i\geq 0$ satisfying 
$\sum_i  M_i = 1$ (POVM).
The probability to find the result $j$ is $P_{j|i}=\tr ( M_j \rho_i )$ if the system  is  in the state
$\rho_i$. The maximal success probability of the receiver reads 
\begin{equation} \label{eq-max_success_proba_POVM}
P_{S}^{\,\rm{opt}} ( \{ \rho_i,\eta_i\}) =\max_{{\rm POVM}\; \{ M_i \} } \sum_{i=1}^{n_A} \eta_i \tr ( M_i \rho_i)\;.
\end{equation}
%

%%%%%%%%%%%%%%%%%%%%%%%%%%%%%%%%%%%%%%%%%%%%%%%%%%%%%%%%%%%%%%%%%%%%%%%%%%%%%%%%%%%%%
\noindent {\bf Theorem 2.} {\it Let  $\rho$ be a state of the bipartite system $AB$
with Hilbert space $\Hh= \Hh_A \otimes \Hh_B$ and let 
$\alpha=\{ \ket{\alpha_i} \}_{i=1}^{n_A}$ be a fixed orthonormal basis
of $\Hh_A$.
Consider the subset $\Cc_A ( \alpha  )\subset \Cc_A$  of all $A$-classical states $\sigma_\Aclass$
such that $\alpha$ is an eigenbasis of $\tr_{B} ( \sigma_\Aclass)$ (\ie $\Cc_A ( \alpha  )$ is the set
of all states $\sigma_\Aclass$ of the form (\ref{eq-A-classical_state}), for arbitrary probabilities
$q_i$ and states $\sigma_{B|i}$ on $\Hh_B$).
Then the maximal fidelity 
$\dss F( \rho,  \Cc_A (\alpha) ) = \max_{\sigma_\Aclass \in \Cc_A (\alpha )} F( \rho, \sigma_\Aclass )$
of $\rho$ to this subset is equal to
\be \label{eq-fidelity_equal_min_error_proba}
F( \rho,  \Cc_A ( \alpha ) )  = 
P_S^{\,\rm{opt\,v.N.}} ( \{ \rho_i,\eta_i \})
\equiv \max_{ \{ \Pi_i \} } \sum_{i=1}^{n_A} \eta_i \tr ( \Pi_i \rho_i)\;,
\ee
where $P_S^{\,\rm{opt\,v.N.}} ( \{ \rho_i,\eta_i \})$ is the maximal success probability over all
von Neumann measurements given by  orthogonal projectors $\Pi_i$ of rank $n_B$ (that is,  
self-adjoint operators on $\Hh$ satisfying $\Pi_i \Pi_j  = \delta_{ij} \Pi_i$ and $\dim ( \Pi_i \Hh )= n_B$), and
\begin{equation} \label{eq-state_Q_discrimination}
\eta_i = \bra{\alpha_i}  \rho_A \ket{\alpha_i} \qquad , \qquad 
\rho_i = \eta_i^{-1} \sqrt{\rho} \ketbra{\alpha_i}{\alpha_i} \otimes 1 \sqrt{\rho}
\end{equation}
(if $\eta_i = 0$ then $\rho_i$ is not defined but does not contribute to the sum in
{\rm (\ref{eq-fidelity_equal_min_error_proba})}).
}
%%%%%%%%%%%%%%%%%%%%%%%%%%%%%%%%%%%%%%%%%%%%%%%%%%%%%%%%%%%%%%%%%%%%%%%%%%%%%%%%%%%%%%%%%%

\vspace{3mm}

This theorem will be proven in Sec.~\ref{sec-derivation_var_formula}.
Note that  the 
$\rho_i$ are quantum states of $AB$ if $\eta_i >0$,
 because the right-hand side of the last identity in (\ref{eq-state_Q_discrimination})
is a non-negative operator and $\eta_i$ is chosen such that $\tr(\rho_i)=1$.  
Moreover, $\{ \eta_i \}_{i=1}^{n_A}$ is a set
of probabilities (since $\eta_i \geq 0$ and $\sum_i \eta_i = \tr (\rho)=1$) and 
$\{ \rho_i,\eta_i\}_{i=1}^{n_A}$ defines a convex decomposition of $\rho$,  \ie
$\rho = \sum_i \eta_i \rho_i$.

Let us assume that $\rho$ is invertible. 
Then the application of a result by Eldar~\cite{Eldar03} shows that 
 the POVM maximizing the success probability $P_S( \{ \rho_i,\eta_i\})$ in (\ref{eq-max_success_proba_POVM}) 
is a von Neumann measurement with projectors $\Pi_i$ of  rank $n_B$, \ie
\be \label{eq-Eldar_result}
F( \rho,  \Cc_A ( \alpha ) )  = P_S^{\,\rm{opt\,v.N.}} ( \{ \rho_i,\eta_i \})
= P_S^{\,\rm{opt}} ( \{ \rho_i,\eta_i \})\quad , \quad \rho >0\,.
\ee
In fact, one may first notice that  all matrices $\rho_i$ have
rank $r_i = n_B$ (for indeed, $\rho_i$  has the same rank as 
$\eta_i \rho^{-1/2} \rho_i = \ketbra{\alpha_i}{\alpha_i} \otimes 1 \sqrt{\rho}$ and 
the latter matrix has rank $n_B$). Next, we argue that 
the $\rho_i$ are linearly independent, in the sense that their eigenvectors $\ket{\xi_{ij}}$ 
form a linearly independent family $\{ \ket{\xi_{ij}} \}_{i=1, \ldots , n_A}^{j=1,\ldots, n_B}$
of vectors in $\Hh$.
Actually, a necessary and sufficient condition for $\ket{\xi_{ij}}$ to be an eigenvector of $\rho_i $
 with eigenvalue $\lambda_{ij}>0$ is 
$\ket{\xi_{ij}}= (\lambda_{ij} \eta_i )^{-1} \sqrt{\rho} \ket{\alpha_i}\otimes \ket{\zeta_{ij}}$,
$\ket{\zeta_{ij}} \in \Hh_B$ being an eigenvector of $R_i= \bra{\alpha_i} \rho \ket{\alpha_i}$
with eigenvalue $\lambda_{ij} \eta_i >0$. 
For any $i$, the  Hermitian invertible matrix
 $R_i$ admits an orthonormal 
eigenbasis $\{ \ket{\zeta_{ij}} \}_{j=1}^{n_B}$. Thanks to the invertibility of $\sqrt{\rho}$,
$\{ \ket{\xi_{ij}} \}_{i=1, \ldots , n_A}^{j=1,\ldots, n_B}$ is a basis of $\Hh$ and thus
the states $\rho_i$ are linearly independent.
It is shown in~\cite{Eldar03} that for such a family of linearly independent states
the second equality in (\ref{eq-Eldar_result}) holds true.

The following result on the Bures-GQD of mixed states is a direct consequence of Theorem~2.

\vspace{2mm}

%%%%%%%%%%%%%%%%%%%%%%%%%%%%%%%%%%%%%%%%%%%%%%%%%%%%%%%%%%%%%%%%%%%%%%%%%%%%%%%%%%%%%%
\noindent {\bf Theorem 3.} {\it For any state $\rho$ of the bipartite system $AB$,
the fidelity to the closest $A$-classical state is given by 
\begin{equation} \label{eq-variationnal_formula}
F_A (\rho ) = \max_{\{ \ket{\alpha_i} \} } \max_{ \{ \Pi_i\}} \sum_{i=1}^{n_A} 
\tr [ \Pi_i \sqrt{\rho} \ketbra{\alpha_i}{\alpha_i} \otimes 1 \, \sqrt{\rho} ] \;,
\end{equation}
where the maxima are over all orthonormal basis $\{ \ket{\alpha_i} \}$ 
of $\Hh_A$ and all orthogonal families $\{ \Pi_i\}_{i=1}^{n_A}$  of  projectors 
of $\Hh_A \otimes \Hh_B$ with rank $n_B$.
Hence, using the notation of theorem 2,
\begin{equation} \label{eq-variationnal_formula_bis}
F_A (\rho) = \max_{\{ \ket{\alpha_i} \} } P_S^{\,\rm{opt\,v.N.}} ( \{ \rho_i,\eta_i \})\,.
\end{equation}
If $\rho>0$ then one can replace  $P_S^{\,\rm{opt\,v.N.}}$ in (\ref{eq-variationnal_formula_bis}) by
the maximal success probability (\ref{eq-max_success_proba_POVM}) over all POVMs.
}
%%%%%%%%%%%%%%%%%%%%%%%%%%%%%%%%%%%%%%%%%%%%%%%%%%%%%%%%%%%%%%%%%%%%%%%%%%%%%%%%%%%%%%%%%

\vspace{3mm}

It is noteworthy to observe that the basis vectors $\ket{\alpha_i}$ can be recovered from 
the states $\rho_i$ and probabilities $\eta_i$
 by forming the  square-root measurement operators  
$M_i  = \eta_i \rho^{-1/2}  \rho_i \rho^{-1/2}$, with $\rho = \sum_i \eta_i \rho_i$ 
(we assume here $\rho>0$).
Actually, such measurement operators are equal to the rank-$n_B$ projectors
$M_i =\ketbra{\alpha_i}{\alpha_i} \otimes 1$.
By bounding from below $P_S^{\,\rm{opt\,v.N.}} ( \{ \rho_i,\eta_i \})$ by
the success probability corresponding to $\Pi_i = M_i$, we obtain
\begin{equation} \label{eq-inequality_square_root_meas}
F_A (\rho) \geq \max_{\{ \ket{\alpha_i} \} } \sum_{i=1}^{n_A} 
 \tr_B \bigl[ \bra{\alpha_i} \sqrt{\rho} \ket{\alpha_i}^2 \bigr]\;.
\end{equation}
The square root measurement plays an important role in the discrimination of almost orthogonal
states~\cite{Hausladen94,Barnum02} and of ensembles of states with certain 
symmetries~\cite{Eldar04,Chou03}.

To illustrate our result, let us study the ambiguous QSD task for some specific states $\rho$. 

\vspace{1mm}

(i) If $\rho$ is an  $A$-classical state, \ie if it admits the decomposition 
(\ref{eq-A-classical_state}), then the basis $\{ \ket{\alpha_i}\}$ maximizing the optimal success probability
in (\ref{eq-variationnal_formula_bis}) coincides
with the basis appearing in this decomposition. With this choice, one obtains $\eta_i = q_i$ and 
$\rho_i = \ketbra{\alpha_i}{\alpha_i} \otimes \sigma_{B|i}$ for all $i$ such that $q_i >0$. 
The states $\rho_i$ are orthogonal and can thus be perfectly discriminated by von Neumann measurements, 
so that 
$F_A(\rho )=P_S^{\,\rm{opt\,v.N.}} ( \{ \rho_i,\eta_i \})=1$.
Reciprocally, if $F_A(\rho)=1$ then $P_S^{\,\rm{opt\,v.N.}}( \{ \rho_i,\eta_i \})=1$ for some
basis $\{ \ket{\alpha_i} \}$ of $\Hh_A$ and  the corresponding $\rho_i$ must be orthogonal,
that is, $\rho_i = \Pi_i \rho_i \Pi_i$ for some orthogonal family $\{ \Pi_i\}$ of   projectors 
with rank  $n_B$.
Hence $\rho = \sum_i \eta_i \rho_i = \sum_i \eta_i \Pi_i \rho_i \Pi_i$, 
$\sqrt{\rho} = \sum_i \Pi_i \sqrt{\rho}\, \Pi_i$, and (\ref{eq-state_Q_discrimination}) entails
$\eta_i \rho_i  =\eta_i \Pi_i \rho_i \Pi_i
 = \sqrt{\rho} \ketbra{\alpha_i}{\alpha_i} \otimes 1 \sqrt{\rho}
= \sqrt{\rho}\, \Pi_i \ketbra{\alpha_i}{\alpha_i} \otimes 1\, \Pi_i \sqrt{\rho}$, implying
$\Pi_i = \ketbra{\alpha_i}{\alpha_i} \otimes 1$ if $\rho$ is invertible. 
Thus $\rho$ is $A$-classical (this was of course to be expected  
since $D_A (\rho )=0$ if and only if $\rho$ is
$A$-classical, see Sec.~\ref{sec-definitions}). 
Therefore, we can interpret our result (\ref{eq-variationnal_formula_bis}) as follows:
{\it the non-zero discord states $\rho$ are such that the states} (\ref{eq-state_Q_discrimination}) 
{\it are non-orthogonal and thus cannot be perfectly discriminated for any orthonormal basis
$\{ \ket{\alpha_i} \}$ of $\Hh_A$.}

\vspace{1mm}

(ii)
If $\rho=\rho_\Psi$ is a pure state, then all $\rho_i$ with $\eta_i >0$ are identical and equal 
to $\rho_\Psi$, so that 
$P_S^{\,\rm{opt}} = P_S^{\,\rm{opt\,v.N.}} = \sup_{\{ \Pi_i\}} \sum_i \eta_i \bra{\Psi} \Pi_i \ket{\Psi} = \eta_{\rm max}$.
One gets back the result $F_A (\rho_\Psi)=\mu_{\rm max}$ of
 Sec.~\ref{sec_pure_states} by optimization over the basis $\{ \ket{\alpha_i} \}$.

\vspace{1mm}

(iii) Let us determine 
the states $\rho$ having the highest possible GQD, \ie  
the smallest possible fidelity $F_A(\rho)$.

\vspace{3mm}

%%%%%%%%%%%%%%%%%%%%%%%%%%%%%%%%%%%%%%%%%%%%%%%%%%%%%%%%%%%%%%%%%%%%%%%%%%%%%%%%%
\noindent {\bf Proposition.} {\it If $n_A \leq n_B$, the smallest fidelity $F_A(\rho)$ for all
states $\rho$ of $AB$ is equal to $1/n_A$. If $r n_A \leq n_B < (r+1) n_A$ with $r=1,2,\ldots$, 
the states $\rho$ with $F_A(\rho)=1/n_A$
are any convex combinations of the $r$ maximally entangled pure states  
$\ket{\Psi_k}= n_A^{-1/2} \sum_{i=1}^{n_A} \ket{\phi_i^{(k)}} \otimes \ket{\psi_i^{(k)}}$, $k=1,\ldots, r$, with
$\braket{\phi_i^{(k)}}{\phi_j^{(k)}}=\delta_{ij}$ and 
$\braket{\psi_i^{(k)}}{\psi_j^{(l)}}=\delta_{kl} \delta_{ij}$. 
}
%%%%%%%%%%%%%%%%%%%%%%%%%%%%%%%%%%%%%%%%%%%%%%%%%%%%%%%%%%%%%%%%%%%%%%%%%%%%%%%%%

\vspace{3mm}

We deduce from this result that the GQD  $D_A(\rho)$ varies between $0$ and $2-2/\sqrt{n_A}$
when  $n_A \leq n_B$. 
By virtue of theorem~1, the proposition, and the inequality $E(\rho) \leq D_A(\rho)$, 
the geometric measure of entanglement $E(\rho)$ also varies between these two values. 
This means that {\it the  most distant states from 
the set of $A$-classical states} $\Cc_A$ {\it are also the most distant from the set of separable states} $\Ss$.
If $n_A \leq n_B < 2 n_A$, these most distant states are always maximally entangled pure states.

\vspace{2mm}

\noindent {\it Proof.}
The success probability  $P_S^{\,\rm{opt\,v.N.}}$ must be clearly larger 
than the highest prior probability $\eta_{\max}  = \max_i \{ \eta_i\}$. 
(A receiver would obtain 
$P_S = \eta_{\rm max}$ by simply guessing that his state is
$\rho_{i_{\rm max}}$, with $\eta_{i_{\rm max}}=\eta_{\max}$; a better strategy is of course to perform the 
von Neumann measurement $\{ \Pi_i \}$ such that  $\Pi_{{i}_{\rm max}}$ projects on 
a $n_B$-dimensional subspace containing the range of $\rho_{i_{\rm max}}$; this range
has a dimension $\leq n_B$  by a similar argument as in the discussion following theorem~2.)
In view of (\ref{eq-variationnal_formula_bis}) and by using $\eta_{\max} \geq 1/n_A$  (since $\sum_i \eta_i = 1$)
we get 
\begin{equation} \label{eq-bound_F_A}
F_A (\rho) \geq \frac{1}{n_A} 
\end{equation}
for any mixed state $\rho$. 

When $n_A \leq n_B$ the bound (\ref{eq-bound_F_A}) is optimum, the value $1/n_A$ being reached for 
the maximally entangled
pure states, see  Sec.~\ref{sec_pure_states}.
Thus $1/n_A$ is the smallest possible fidelity.
Let $\rho$ be a state having such a fidelity $F_A (\rho) = 1/n_A$. 
According to (\ref{eq-variationnal_formula_bis}) and since it has been argued before that
$P_S^{\,\rm{opt\,v.N.}} \geq \eta_{\rm max} \geq 1/n_A$, $F_A (\rho) = 1/n_A$ implies
that $P_S^{\,\rm{opt\,v.N.}} ( \{ \rho_i,\eta_i \})=1/n_A$  whatever the orthonormal basis $\{ \ket{\alpha_i} \}$.
It is intuitively clear that this can only happen if
the receiver gets a collection of identical states $\rho_i$ with equal prior 
probabilities $\eta_i = 1/n_A$. A rigorous proof of
this fact is given in the appendix. From (\ref{eq-state_Q_discrimination})
and $\rho= \sum \eta_i \rho_i$ we then obtain  $\bra{\alpha_i} \rho_A \ket{\alpha_i}=1/n_A$ 
and $\rho_i = \rho$ for any $i=1,\ldots , n_A$ and any orthonormal basis $\{ \ket{\alpha_i} \}$.
The first equality implies $\rho_A = 1/n_A$. By replacing the spectral decomposition 
$\rho = \sum_{k} p_k \ketbra{\Psi_k}{\Psi_k}$ into  (\ref{eq-state_Q_discrimination}),
the second equality yields
$\tr_B ( \ketbra{\Psi_k}{\Psi_l} )= n_A^{-1} \delta_{kl}$
for all $k$, $l$ with $p_k p_l \not= 0$.
Taking advantage of this identity for $k=l$, one finds that the eigenvectors
 $\ket{\Psi_k}$ of $\rho$ with positive eigenvalues $p_k$ have all their Schmidt eigenvalues  equal to  $1/n_A$,
that is, their Schmidt decompositions read
$\ket{\Psi_k} = n_A^{-1/2} \sum_{i=1}^{n_A} \ket{\phi_i^{(k)}} \otimes \ket{\psi_i^{(k)}}$. 
Moreover,
$\tr_B ( \ketbra{\Psi_k}{\Psi_l} )=0$ is equivalent to 
$\Vv_B^{(k)} \bot \Vv^{(l)}_B$ with $\Vv_B^{(k)} = \Span \{ \ket{\psi_i^{(k)}}  \}_{i=1}^{n_A} \subset \Hh_B$.
If $n_B < (r+1) n_A$ then at most $r$ subspaces $\Vv_B^{(k)} $ may
be pairwise orthogonal.  Thus at most $r$ eigenvalues $p_k$ are non-zero. 
\finpro

\vspace{2mm}

Let us now discuss the case $n_A > n_B$.
In that case the smallest value of the maximal fidelity $F_\Ss(\rho)$ to a separable state  
is equal to $1/n_B$ and $F_\Ss(\rho)=1/n_B$ when $\rho$ is a  pure maximally entangled state. 
This is a consequence of
 (\ref{eq-fidelity_as_convex_roof}) and of the bound
$F_\Ss (\rho_\Psi)  \geq 1/n_B$ for pure states $\rho_\Psi$ (see Sec.~\ref{sec_pure_states}). 
As a result, the geometric measure of entanglement $E(\rho)$ varies between
$0$ and  $2-2/\sqrt{n}$ with $n=\min\{ n_A,n_B\}$, in both cases $n_A\leq n_B$ and 
$n_B > n_A$. We could not establish a similar result
for the GQD $D_A(\rho)$. When $n_A > n_B$,
the bound (\ref{eq-bound_F_A}) is still correct but it is not optimal, \ie 
there are no states $\rho$ with fidelities $F_A(\rho)$ equal to $1/n_A$. 
Indeed, following the same lines as in the proof above, one shows that if $F_A(\rho)=1/n_A$ then 
the eigenvectors $\ket{\Psi_k}$ of $\rho$ with non-zero eigenvalues
must have maximally mixed marginals $[\rho_{\Psi_k}]_A = 1/n_A$. But this  is impossible 
since $\rank ( [\rho_{\Psi_k}]_A) \leq n_B$  by (\ref{eq-Schmidt_decomposition}).
According to the results of Sect.~\ref{sec_pure_states},  pure states  $\rho_\Psi$
have fidelities $F_A(\rho_\Psi) \geq 1/n_B$, so one may expect that 
states close enough to pure states have fidelities  close to $1/n_B$ or larger. 
This can be shown rigorously by invoking the bound 
\begin{equation} \label{eq-refined_bound_n_A>n_B}
F_A (\rho) \geq \frac{\| \rho\|}{n_B} + \frac{1-\|\rho\|}{n_A} \frac{n_B-\delta_\rho}{n_B}
\end{equation}
where $\| \rho\|$ is the norm of $\rho$ and $\delta_\rho = 0$ if $\rank (\rho) \leq n_B$ and $1$ otherwise.
This bound can be established as follows.
Let us write $\rho = p \ketbra{\Psi}{\Psi} + (1-p) {\rho}'$ where $\ket{\Psi}$ is 
the eigenvector of $\rho$ with  maximal eigenvalue $p = \|\rho\|$ and the density matrix
$\rho'$ has support on $[\complex \ket{\Psi} ]^\bot$. 
Choosing an orthonormal family $\{ \Pi_i\}$
of projectors of rank $n_B$ satisfying $\Pi_1 \ket{\Psi_1} = \ket{\Psi_1}$, we get 
from (\ref{eq-variationnal_formula_bis})
\begin{equation} \label{eq-derivation_refined_bound_n_A>n_B}
F_A ( \rho ) \geq \sum_i \eta_i \tr ( \Pi_i \rho_i ) = p \bra{\alpha_1} \tr_B ( \ketbra{\Psi}{\Psi} ) \ket{\alpha_1}
+ (1-p) \sum_i {\eta}_i' \tr ( \Pi_i {\rho}_i'  )
\end{equation}
with ${\eta}_i' {\rho}_i' = \sqrt{{\rho}'} \ketbra{\alpha_i}{\alpha_i} \otimes 1   \sqrt{\rho'}$
and ${\eta}_i' = \bra{\alpha_i} {\rho}_A' \ket{\alpha_i}$.
Let us fix the orthonormal basis $\{ \ket{\alpha_i} \}$ such that $\ket{\alpha_1}$ is the eigenvector 
with maximal eigenvalue $\mu_{\rm max}$ in the Schmidt decomposition (\ref{eq-Schmidt_decomposition})
of $\ket{\Psi}$.
This leads to the maximal possible value $p \mu_{\rm max}$ of 
the first term in the right-hand side of (\ref{eq-derivation_refined_bound_n_A>n_B}).
We now bound the sum in this right-hand side
by  its $i_{m}\,$th term ${\eta}_{\rm max}' \tr ( \Pi_{i_{m}} \rho_{i_{m}}')$, where
$i_{m}$ is the index $i$ such that ${\eta}_i'$ is maximum, \ie $\eta_{i_m}' = {\eta}_{\rm max}'$. 
If $i_{m} >1$, one can find orthogonal projectors $\Pi_1$ and $\Pi_{i_{m}}$ such that
$\ket{\Psi} \in \Pi_1 {\cal H}$ and 
${\rho}_{i_{m}}' \Hh \subset \Pi_{i_{m}} {\cal H} \subset [\complex \ket{\Psi} ]^\bot$
(recall that the ${\rho}_{i}'$ have ranks $\leq n_B$).
If $i_{m} =1$, we choose $\Pi_1 = \ketbra{\Psi}{\Psi} + \Pi_1'$ where $\Pi_1'$ is the spectral projector
of ${\rho}_{1}'$ associated to the $(n_B-1)$ highest eigenvalues 
${q}_1' \geq {q}_2' \geq \cdots \geq {q}_{n_B-1}'$.
In all cases, $\tr ( \Pi_{i_{m}} \rho_{i_{m}}') \geq 1 - {q}_{n_B}'$. If $\rank (\rho) \leq n_B$ then 
$q_{n_B}'=0$, otherwise we bound $q_{n_B}'$ by $1/n_B$ 
(since $\sum_{j=1}^{n_B} {q}_j' =1$).
Collecting the above results and using the inequalities $\mu_{\rm max} \geq 1/n_B$ and 
${\eta}_{\rm max}' \geq 1/n_A$ (since $\sum_{i=1}^{n_A} \eta_i'= 1$), one gets 
(\ref{eq-refined_bound_n_A>n_B}). Note that this bound is stronger than 
(\ref{eq-bound_F_A}) only for states $\rho$ satisfying $\| \rho\| \geq (1+n_A-n_B)^{-1}$ 
or $\rank (\rho) \leq n_B$. 
In summary, we can only conclude from the analysis above that  when $n_A > n_B$
the smallest possible fidelity $\min_{\rho \in {\cal E}} F_A(\rho)$  
 lies in the interval $( 1/n_A, 1/n_B]$.

%%%%%%%%%%%%%%%%%%%%%%%%%%%%%%%%%%%%%%%%%%%%%%%%%%%%%%%%%%%%%%%%%%%%%
\subsection{Derivation of the variational formula (\ref{eq-variationnal_formula})} 
\label{sec-derivation_var_formula}
%%%%%%%%%%%%%%%%%%%%%%%%%%%%%%%%%%%%%%%%%%%%%%%%%%%%%%%%%%%%%%%%%%%%%

To prove theorems 2 and 3, we start
by evaluating the trace norm in (\ref{eq-fidelity}) by means of the formula
$\| T \|_1 = \max_{U} | \tr ( U T) |$, the maximum being over all unitaries on
$\Hh$. By (\ref{eq-A-classical_state_bis}),  
\begin{eqnarray} \label{eq-using_formula_for_trace_norm}
\nn
\sqrt{F (\rho, \sigma_\Aclass ) } 
& = &
\max_U \bigl| \tr ( U \sqrt{\rho} \sqrt{\sigma_\Aclass} ) \bigr| 
\\
\nn
& = & 
\max_U \Bigl| \sum_{i,j} \sqrt{q_{ij}} \tr ( U \sqrt{\rho}\, \ketbra{\alpha_i}{\alpha_i} \otimes 
 \ketbra{\beta_{j|i}}{\beta_{j|i}} ) \Bigr|
\\
\nn
& = & 
\max_{\{ \ket{\Phi_{ij}} \}}  \Bigl| \sum_{i,j} \sqrt{q_{ij}} \bra{\Phi_{ij}}
  \sqrt{\rho} \ket{\alpha_i \otimes \beta_{j|i}} \Bigr|
\\
& = & 
 \max_{\{ \ket{\Phi_{ij}} \} }  \sum_{i,j} \sqrt{q_{ij}} \bigl| \bra{\Phi_{ij}}
  \sqrt{\rho} \ket{\alpha_i \otimes \beta_{j|i}} \bigr| \;.
\end{eqnarray}
In the third line we have replaced the maximum over unitaries $U$ by a maximum 
over all orthonormal basis $\{ \ket{\Phi_{ij}} \}$ of $\Hh$ 
(with $\ket{\Phi_{ij}}=U^\dagger \ket{\alpha_i \otimes \beta_{j|i}}$).  The last equality 
in (\ref{eq-using_formula_for_trace_norm}) can be explained as follows.
The expression in the last line is clearly larger than that of the third line; since for any $i$ and $j$
one can choose the phase factors of the  vectors $\ket{\Phi_{ij}}$ in such a way
that $ \bra{\Phi_{ij}} \sqrt{\rho} \ket{\alpha_i \otimes \beta_{j|i}} \geq 0$, the two expressions are
in fact equal. 

One has to maximize the last member of (\ref{eq-using_formula_for_trace_norm}) 
over all families of $i$-dependent orthonormal basis  
$\{ \ket{\beta_{j|i}} \}$ of $\Hh_B$ and all probabilities $q_{ij}$.
The maximum over the probabilities $q_{ij}$ is easy to evaluate 
by using  the Cauchy-Schwarz inequality and $\sum_{i,j} q_{ij} =1$. It is reached
for 
\begin{equation} \label{eq-optimal_proba}
q_{ij} 
= \frac{ |   \bra{\Phi_{ij}}  \sqrt{\rho} \ket{\alpha_i \otimes \beta_{j|i}} |^2}{\sum_{ij}  | \bra{\Phi_{ij}}
  \sqrt{\rho} \ket{\alpha_i \otimes \beta_{j|i}} |^2} \;.
\end{equation}
We thus obtain
\begin{equation}\label{eq-max_fidelity1}
F (\rho, \Cc_A(\alpha) ) 
=
\max_{ \{ \ket{\beta_{j|i}} \} } \max_{ \{ q_{ij} \} } F (\rho, \sigma_\Aclass)
 = 
\max_{ \{ \ket{\beta_{j|i}} \}} \max_{\{ \ket{\Phi_{ij}} \} } 
 \sum_{i,j}  \bigl| \braket{\psi_{j|i}}{\beta_{j|i}} \bigr|^2 
\end{equation}  
where we have set  $\ket{\psi_{j|i}} = \bra{\alpha_i} \sqrt{\rho} \ket{\Phi_{ij}} \in \Hh_B$.
We proceed to evaluate the maximum over $\{ \ket{\beta_{j|i}} \}$ and $\{ \ket{\Phi_{ij}} \}$. Let us 
fix $i$ and consider the orthogonal family of projectors of $\Hh$ of rank $n_B$ defined by
\begin{equation} \label{eq-def_Pi_i_projectors}
\Pi_i = \sum_{j} \ketbra{\Phi_{ij}}{\Phi_{ij}}\;.
\end{equation}
By the Cauchy-Schwarz inequality, for any fixed $i$ one has
\begin{equation} \label{eq-sup_over_chi_j}
\max_{ \{ \ket{\beta_{j|i}} \}} \sum_{j} \bigl| \braket{\psi_{j|i}}{\beta_{j|i}} \bigr|^2 
\leq
\sum_{j} \| \psi_{j|i} \|^2 
 = \tr \bigl[ \Pi_i \sqrt{\rho} \ketbra{\alpha_i}{\alpha_i} \otimes 1 \, \sqrt{\rho} \bigr] \;.
\end{equation}
Note that (\ref{eq-sup_over_chi_j}) is an inequality 
if the vectors $\ket{\psi_{j|i}}$ are orthogonal for different $j$'s.
We now show that this is the case provided that the $\ket{\Phi_{ij}}$ are chosen appropriately.
In fact, let us take an arbitrary orthonormal basis $\{ \ket{\Phi_{ij}} \}$ of $\Hh$ and 
consider  the Hermitian 
$n_B \times n_B$ matrix $S^{(i)}$ with coefficients given by the scalar
products 
$S^{(i)}_{jk} = \braket{\psi_{j|i}}{\psi_{k|i}}$.
One can find a unitary matrix $V^{(i)}$ such that 
$\widetilde{S}^{(i)} = (V^{(i)} )^\dagger S^{(i)} V^{(i)}$ is diagonal 
and has non-zero diagonal elements in the first $r_i$ raws,
where $r_i$ is the rank of $S^{(i)}$. Let 
$\ket{\widetilde{\Phi}_{ij}} = \sum_{l=1}^{n_B} V^{(i)}_{lj} \ket{\Phi_{il}}$. Then
$\{ \ket{\widetilde{\Phi}_{ij}}\}$ is an orthonormal basis of $\Hh$ and
$\sum_{j} \ketbra{\widetilde{\Phi}_{ij}}{\widetilde{\Phi}_{ij}} = \Pi_i$.
Moreover, the vectors 
$\ket{\widetilde{\psi}_{j|i}} = \bra{\alpha_i} \sqrt{\rho} \ket{\widetilde{\Phi}_{ij}}
 = \sum_{l=1}^{n_B} V_{lj}^{(i)} \ket{\psi_{l|i}}$
form an orthogonal set $\{ \ket{\widetilde{\psi}_{j|i}} \}_{j=1}^{r_i}$ and  
vanish for $j>r_i$. Therefore, for any fixed orthogonal family $\{ \Pi_i \}_{i=1}^{n_A}$ of  projectors
of rank $n_B$, there exists an orthonormal basis $\{ \ket{{\Phi}_{ij}}\}$ of 
$\Hh$ such that (\ref{eq-def_Pi_i_projectors}) holds and the inequality in
(\ref{eq-sup_over_chi_j}) is an equality. Substituting this equality into (\ref{eq-max_fidelity1}), one
finds
\begin{equation} \label{eq-I_was_stupid}
F (\rho, \Cc_A(\alpha) ) 
= \max_{ \{ \Pi_i \}} \sum_{i,j}  \| \widetilde{\psi}_{j|i} \|^2 
=  \max_{ \{ \Pi_i \}} \sum_{i} \tr [ \Pi_i \sqrt{\rho} \ketbra{\alpha_i}{\alpha_i} \otimes 1\,\sqrt{\rho} ]\;,
\end{equation}
which yields the result (\ref{eq-fidelity_equal_min_error_proba}).
The formula (\ref{eq-variationnal_formula}) is obtained by maximization 
over the basis $\{ \ket{\alpha_i} \}$.
\finpro

%%%%%%%%%%%%%%%%%%%%%%%%%%%%%%%%%%%%%%%%%%%%%%%%%%%%%%%%%%%%%%%%%%%%%
\subsection{Closest $A$-classical states}
%%%%%%%%%%%%%%%%%%%%%%%%%%%%%%%%%%%%%%%%%%%%%%%%%%%%%%%%%%%%%%%%%%%%%

The proof of the previous subsection
also  gives an algorithm to find the closest $A$-classical states to a given mixed
state $\rho$. To this end, one must find  the orthonormal 
basis  $\{ \ket{\alpha_i^{\rm{opt}}} \}$ of $\Hh_A$ maximizing  
$P_S^{\,\rm{opt\,v.N.}} ( \{ \rho_i,\eta_i \})$ in (\ref{eq-variationnal_formula_bis})
and the optimal von Neumann measurement $\{ \Pi_i^{\rm{opt}} \}$ yielding the minimal error in the 
discrimination of the ensemble $\{ \rho_i^\opt, \eta_i^\opt\}$ associated to  
$\{ \ket{\alpha_i^{\rm{opt}}}\}$ in Eq. (\ref{eq-state_Q_discrimination}).

\vspace{2mm}

\noindent {\bf{Theorem 4.}}
{\it  
The closest $A$-classical states to $\rho$ are
\begin{equation} \label{eq-again_I_was_stupid}
\sigma_\rho = \frac{1}{F_A(\rho)} 
\sum_{i=1}^{n_A} \ketbra{\alpha_i^{\rm{opt}}}{\alpha_i^{\rm{opt}}} 
\otimes \bra{\alpha_i^{\rm{opt}}} \sqrt{\rho}\, \Pi_i^{\rm{opt}} \sqrt{\rho} \ket{\alpha_i^{\rm{opt}}}  \;,
\end{equation}
where $\{ \ket{\alpha_i^{\rm{opt}}} \}_{i=1}^{n_A}$ and 
$\{ \Pi_i^{\rm{opt}} \}_{i=1}^{n_A}$ are such that
$F_A(\rho)=\sum_{i} \tr [ \Pi_i^{\rm{opt}}  
\sqrt{\rho} \ketbra{\alpha_i^{\rm{opt}}}{\alpha_i^{\rm{opt}}} \otimes 1\, \sqrt{\rho} ]$
(see Eq.{\rm (\ref{eq-variationnal_formula})}).
}

\vspace{3mm}

\noindent {\it Proof.} This  follows directly from the proof of the previous subsection.
Actually, by using the  expression (\ref{eq-optimal_proba}) of the optimal probabilities $q_{ij}$
and the fact that the  Cauchy-Schwarz inequality (\ref{eq-sup_over_chi_j}) is an equality
if and only if $\ket{\beta_{j|i}} = \ket{\psi_{j|i}}/\| \psi_{j|i} \|$ when $\| \psi_{j|i} \|\not= 0$,
we conclude that the closest $A$-classical states to $\rho$ are given  by 
(\ref{eq-A-classical_state}) with $\ket{\alpha_i}= \ket{\alpha_i^\opt}$ and 
\begin{equation}
q_i \sigma_{B|i} = 
\sum_{j=1}^{n_B} q_{ij} \ketbra{\beta_{j|i}}{\beta_{j|i}}
=
\frac{\sum_{j=1}^{n_B} \ketbra{\widetilde{\psi}_{j|i}}{\widetilde{\psi}_{j|i}}}
{\sum_{i=1}^{n_A} \sum_{j=1}^{n_B} \| \widetilde{\psi}_{j|i} \|^2}   \;.
\end{equation}
The denominator is equal to $F_A (\rho)$, see (\ref{eq-I_was_stupid}). The numerator
is the same as the second factor in the right-hand side of (\ref{eq-again_I_was_stupid}). For indeed,  
by construction
$\ket{\widetilde{\psi}_{j|i}} = \bra{\alpha_i^{\rm{opt}}} \sqrt{\rho} \ket{\widetilde{\Phi}_{ij}}$
and $\sum_{j} \ketbra{\widetilde{\Phi}_{ij}}{\widetilde{\Phi}_{ij}} = \Pi_i^{\rm{opt}}$.
\finpro

\vspace{2mm}

   Let us stress that the optimal measurement $\{ \Pi_i^\opt\}$ and basis 
$\{ \ket{\alpha_i^\opt}\}$
may not be unique, so that $\rho$ may have several closest $A$-classical  states $\sigma_\rho$.
This is the case for instance when $\rho$ is a pure state with a degenerate 
maximal Schmidt eigenvalue, as we have seen in theorem~1.  
If $\sigma_\rho= \sum q_i \ketbra{\alpha_i^\opt}{\alpha_i^\opt} \otimes \sigma_{B|i}$ and
$\sigma_\rho'= \sum q_i' \ketbra{\alpha_i^\opt}{\alpha_i^\opt} \otimes \sigma_{B|i}'$ are two
closest $A$-classical states to $\rho$ with the same basis $\{ \ket{\alpha_i^\opt} \}$ 
then so are all convex combinations
$\sigma_\rho (\eta) = \eta \sigma_\rho + (1-\eta) \sigma_\rho'$ with $0 \leq \eta \leq 1$.
This fact  is a direct consequence of the convexity of the square Bures distance
(property (i) in Section~\ref{sec-definitions}), given that $\sigma_\rho(\eta) \in \Cc_A$.  
As a result, states $\rho$ having more than one closest $A$-classical state 
will generally admit a continuous family of such states.

%%%%%%%%%%%%%%%%%%%%%%%%%%%%%%%%%%%%%%%%%%%%%%%%%%%%%%%%%%%%%%%%%%
\section{Conclusions}
%%%%%%%%%%%%%%%%%%%%%%%%%%%%%%%%%%%%%%%%%%%%%%%%%%%%%%%%%%%%%%%%%%%

We have established in this paper a link between  ambiguous 
QSD and the problem of finding the minimal Bures distance
of a state $\rho$ of a bipartite system $AB$ to a state with vanishing quantum discord. 
More precisely, the maximal fidelity between $\rho$ and an
$A$-classical (\ie zero-discord) state coincides
with the maximal success probability in discriminating the $n_A$ states $\rho_i^\opt$ with prior
probabilities $\eta_i^\opt$ given by Eq.(\ref{eq-state_Q_discrimination}),   
$n_A$ being the space dimension of subsystem $A$ (theorem~3). These states and probabilities depend upon an optimal
orthonormal basis $\{ \ket{\alpha_i^\opt} \}$ of $A$.
The closest $A$-classical states to $\rho$ are, in turn, given in terms of this optimal
basis and of the optimal von Neumann measurements in the discrimination of
 $\{ \rho_i^\opt , \eta_i^\opt \}$ (theorem~4).    Finally, we have shown that when $n_A \leq n_B$, 
the ``most quantum'' states characterized by the highest possible 
distance to the set of $A$-classical states are the maximally entangled pure states,
or convex combinations of such states with reduced $B$-states having supports on orthogonal subspaces.
These states are also the most distant from the set of separable states. 

   As stated in the introduction,
the QSD task can be solved for $n_A=2$ states. Thus the aforementioned results
provide a method to find  the geometric discord $D_A$ and the closest $A$-classical states for
 bipartite systems composed of a qubit $A$ and a subsystem $B$ with 
arbitrary space dimension $n_B \geq  2$.
In particular,  explicit formulae  can be derived for two qubits in states
with maximally mixed marginals and for 
$(n_B+1)$-qubits in the DQC1 algorithm~\cite{companion_paper}.
For subsystems $A$ with higher space dimensions $n_A > 2$, several open issues deserve further
studies.
Firstly, it would be desirable to characterize  
the ``most quantum'' states when $n_A> n_B$.
Secondly, it is not excluded that the specific QSD task 
associated to the minimal Bures distance admits an explicit solution.
Thirdly, the relation of $D_A$ with the geometric measure of entanglement 
in tripartite systems  should
be investigated; in particular, there may exist some inequality analogous to  
the monogamy relation~\cite{Koashi04} between the quantum discord and the entanglement of formation.  

Let us emphasize that our results may shed new light on
dissipative dynamical processes involving decoherence, \ie evolutions towards 
classical states. In fact, our analysis may allow in some cases to determine the geodesic segment 
linking a given state $\rho_0$ with non-zero
discord to its closest $A$-classical state $\sigma_{\rho_0}$. Such a piece of geodesic is contained in the set
of all states $\rho$ having the same closest $A$-classical state $\sigma_\rho = \sigma_{\rho_0}$ as $\rho_0$.
It would be of interest to compare   in specific physical examples
the Bures geodesics  with the actual paths followed by the 
density matrix during the dynamical evolution.

\vspace{5mm}

%%%%%%%%%%%%%%%%%%%%%%%%%%%%%%%%%%%%%%%%%%%%%%%%%%%%%%%%%%%%%%%%%%%%%%%%%%%%%%%%%%%%%%%%

\noindent {\bf Acknowledgements:}
We are grateful to Maria de los Angeles Gallego for revising the derivation of 
Eq.(\ref{eq-variationnal_formula}) and for
providing us some notes about the calculation of the geometric discord for Werner states. 
D.S. thanks Aldo Delgado for interesting discussions.
We acknowledge financial support from
the French project no. ANR-09-BLAN-0098-01, 
the Chilean Fondecyt project no. 100039, and the project Conicyt-PIA anillo no. ACT-1112 
``Red de An\'alisis estoc\'astico y aplicaciones''.

\vspace{5mm}

\appendix
\renewcommand{\theequation}{\Alph{section}\arabic{equation}}
\setcounter{equation}{0}
%%%%%%%%%%%%%%%%%%%%%%%%%%%%%%%%%%%%%%%%%%%%%%%%%%%%%%%%%%%%%%%%%%%%
\section{Necessary and sufficient condition 
for the optimal success probability to be equal to the inverse number of states} \label{app-C}
%%%%%%%%%%%%%%%%%%%%%%%%%%%%%%%%%%%%%%%%%%%%%%%%%%%%%%%%%%%%%%%%%%%%%%%%%%

Let $\{ \rho_i\}_{i=1}^{n_A}$ be a family of $n_A$ states on $\Hh$ 
with prior probabilities $\eta_i$, where $n_A=N/n_B$ is a divisor of 
$\dim (\Hh)=N$. We assume that the $\rho_i$ have ranks $\rank (\rho_i) \leq n_B$ for any $i$.
Let $P_S^{\,\rm{opt\,v.N.}} ( \{ \rho_i,\eta_i \} )$ be the optimal success probability in discriminating 
the states $\rho_i$, defined by Eq.(\ref{eq-fidelity_equal_min_error_proba}).
We prove in this appendix that
$P_S^{\,\rm{opt\,v.N.}} ( \{ \rho_i,\eta_i \} )=1/n_A$ if and only if 
$\eta_i=1/n_A$  for any $i$ and all states $\rho_i$ are identical.

The conditions $\eta_i=1/n_A$ and $\rho_i=\rho$
are clearly sufficient to have  $P_S^{\,\rm{opt\,v.N.}} ( \{ \rho_i,\eta_i \} )=1/n_A$ 
(a measurement cannot distinguish the identical states $\rho_i$ and thus cannot do better
than a random choice with equal probabilities).
We need to show that they are also necessary conditions. Let us assume 
$P_S^{\,\rm{opt\,v.N.}} ( \{ \rho_i,\eta_i \} )=1/n_A$. 
The equality $\eta_i=1/n_A$ for all $i$ is obvious from the bounds
$P_S^{\,\rm{opt\,v.N.}} ( \{ \rho_i,\eta_i \}) \geq \eta_{\rm max}\equiv \max_i \{ \eta_i \}$ and 
$\eta_{\max} \geq 1/n_A$ (see Sec.~\ref{eq-main_result}). 
Therefore, according to our hypothesis, any orthogonal family $\{ \Pi_i\}_{i=1}^{n_B}$ of  projectors of rank $n_B$
satisfies $ \sum_i \tr ( \Pi_i \rho_i ) = n_A P_S ( \{ \rho_i,\eta_i \}) \leq 1$.
We now argue that the states $\rho_i$ have ranges contained in a  common subspace
$\Vv$.
In fact, let  $\Vv$ be the $n_B$-dimensional subspace of $\Hh$ spanned  by the eigenvectors
of $\rho_1$ associated to the $n_B$ highest  eigenvalues (including degeneracies), and let us denote by 
$\Pi_1$ the projector onto $\Vv$.  Then
$\rho_1 \Hh \subset \Vv$ (since we have assumed $\rank (\rho_1) \leq n_B$) and thus
$\rho_1 = \Pi_1 \rho_1$. Thanks to the inequality above,
$1 \geq   \sum_i \tr ( \Pi_i \rho_i ) \geq \tr (\Pi_1 \rho_1)=1$.
It follows that
$\tr (\Pi_2 \rho_2)=0$ for any  projector $\Pi_2$ of rank $n_B$ orthogonal to $\Pi_1$.
Hence $\rho_2$, and similarly all $\rho_i$, $i=3,\ldots , n_A$, have ranges 
contained in $\Vv$. This proves the aforementioned claim.
 
In order to show that all the states $\rho_i$ are identical, we further
introduce, for each $1 \leq k \leq n_B$, some  $n_B$-dimensional subspace 
$\Vv^{(k)}$  containing the eigenvectors  associated to the $k$ highest eigenvalues 
$\lambda_1 \geq \ldots \geq  \lambda_k$ of $\rho_1$,  the other eigenvectors being orthogonal to $\Vv^{(k)}$
(then $\Vv^{(n_B)}=\Vv$). 
We also choose a $n_B$-dimensional subspace $\Ww^{(k)} \subset \Hh$ orthogonal to $\Vv^{(k)}$ such that 
$\Ww^{(k)} \oplus \Vv^{(k)} \supset \Vv$. Let 
$\{ \Pi_i ^{(k)}\}_{i=1}^{n_A}$ be an orthogonal family of projectors of rank $n_B$ such that
$\Pi_1^{(k)}$ and $\Pi_2^{(k)}$ are the projectors onto $\Vv^{(k)}$ and $\Ww^{(k)}$,
respectively.  Then  
\begin{equation} \label{eq-appendix}
1 \geq  \sum_i \tr ( \Pi_i^{(k)} \rho_i ) 
= \tr ( \Pi_1^{(k)} \rho_1 ) + \tr [ (1- \Pi_1^{(k)}) \rho_2 ]
 =  1 +  \lambda_1 + \cdots + \lambda_k - \tr ( \Pi_1^{(k)} \rho_2 )\;,
\end{equation}
where we used $\sum_{i} \Pi_i^{(k)}=1$ and $\rho_i \Hh  \subset \Ww^{(k)} \oplus \Vv^{(k)}$ in the
first equality. 
By virtue of the min-max theorem, $\tr ( \Pi_1^{(k)} \rho_2 )$ is smaller than the sum of 
the $k$ highest eigenvalues of $\rho_2$ (including degeneracies). By (\ref{eq-appendix}), 
this sum is larger  than the sum $\lambda_1 + \cdots + \lambda_k$ of the $k$ highest eigenvalues of $\rho_1$.
By exchanging the roles of $\rho_1$ and $\rho_2$, we obtain the reverse equality.
Since moreover $k$ is arbitrary between $1$ and $n_B$, it follows that
$\rho_1$ and $\rho_2$ have identical eigenvalues. By using (\ref{eq-appendix}) again,
$\tr ( \Pi_1^{(k)} \rho_2) $ is equal to the sum of the $k$ highest eigenvalues of $\rho_2$.
Hence the $k$ corresponding eigenvectors of $\rho_2$ are contained in the $k$-dimensional 
subspace $\Vv^{(k)} \cap \Vv$. 
Since $k$ is arbitrary, this proves that $\rho_1$ and $\rho_2$ 
have identical eigenspaces.  Therefore $\rho_1 = \rho_2$. Repeating the same argument
for the other states $\rho_i$, $i \geq 3$, we obtain $\rho_1=\cdots = \rho_{n_A}$.

%%%%%%%%%%%%%%%%%%%%%%%%%%%%%%%%%%%%%%%%%%%%%%%%%%%%%%%%%%%%%%%%%%%%%%%%%%%%%%%%%%%%%%%%%%

\end{document}